# CONet: A Cognitive Ocean Network

Huimin Lu

**Abstract**

The scientific and technological revolution of the Internet of Things has begun in oceanography. Historically, humans have observed the ocean from outside the ocean to study it. In recent years, changes have been made to the interior of the ocean, and laboratories have been built on the seafloor. Approximately 70.8% of the Earth is covered by ocean and rivers. The Ocean of Things is expected to be important for disaster prevention, ocean resource exploration, and underwater environmental monitoring. Different from traditional wireless sensor networks, the Ocean of Things has its own unique features, such as low reliability and narrow bandwidth. These features may be great challenges for the Ocean of Things. Furthermore, the integration of the Ocean of Things with artificial intelligence has become a topic of increasing interest for oceanology research fields. The Cognitive Ocean Network (CONet) will become the mainstream of future ocean science and engineering development. In this paper, we define CONet. The contributions of our paper are as follows: (1) a CONet architecture is proposed and described in detail; (2) important and useful demo applications of CONet are proposed; and (3) future trends in CONet research are presented.

## 1. Introduction

The Internet of Things, artificial intelligence, robotics, and blockchains are core technologies of "Industry 4.0," which is to build and realize a super smart society [1]. However, most research is focused on traditional territorial wireless sensor networks. As far as we known, there are few studies on the ocean of things. In 2016, the U.S. National Science Foundation officially announced that the world's largest ocean observation program, Ocean Observation Initiative, had been completed and put into use. It has more than 900 probes in seven major systems that distribute real-time observations from the seabed to the world, and live 15-minute deep-sea hydrothermal activities are reported every three hours. In 2015, the world's largest submarine observation network, S-Net, was built in Japan. From the east coast of Honshu Island to Japanese trenches with depths of 8000 meters, 150 seismic monitoring stations were set up, and 5,700 kilometers of submarine networks were connected by cables. Additionally, the EU EMSO program has increased the number of ocean host stations to 15, and 50 research institutions from 14 countries have participated. In this section, we overview all of these underwater networks.

### 1.1 Ocean Observation Initiative (OOI)

The United States was the first country in the world to make scientific observations using the Ocean of Things, and after more than ten years of research, a nationwide subsea scientific observation network plan was built. In 1995, under the auspices of the NSF, the United States established the first National Ocean Observation Committee to plan the United States Marine Observation Network.

There are various observational tasks of the OOI observation network. First, it continuously provides power and transfers data through optical cables. Second, the anchor connection method is used, wherein the instrument battery on the experimental platform is autonomously powered and satellites transmit observation data to the land information-processing center. Third, the mobile observation method is used, wherein an underwater glider or underwater robots provide large-scale and high-resolution multitemporal observations.

### 1.2 NEPTUNE Canada

NEPTUNE and VENUS are sister projects of Victoria University, Canada, both of which are advanced optical cable observation systems and have many similar observation concepts. The underwater observation range of NEPTUNE is 17–2660 m. The whole system is based on an 800-km submarine fiber cable. The cable starts at west Vancouver Island, passes through the continental shelf into the abyssal plain, and extends outward to the extended center of the mid-ocean ridge, and finally forms a loop.

### 1.3 EMSODEV

The European Multidisciplinary Seafloor Observatory Development of Instrument Module (EMSODEV) project is a subproject of the EU Horizon 2020 program. Its overall objective is to develop, test and deploy the EMSO universal instrument module, EGIM, and promote the full use and operation of the EMSO-dispatched research infrastructure. This module will provide accurate, consistent, and comparable long-term measurements of oceanographic parameters that are key to addressing urgent social and scientific challenges, such as climate change, marine ecosystem disturbances, and marine disasters.

### 1.4 DONET/S-Net

DONET has been supported by Japan's Ministry of Education since 2006 and was implemented by JAMSTEC, a marine science and technology research institute in Japan. It was completed in 2011. As Japan is an earthquake-prone country, early warning of earthquakes and tsunamis has been the primary purpose of DONET's submarine observations. The Japanese Islands are located at the intersection of the Eurasian Plate, Pacific Plate, Philippine Plate, and North American Plate. The earthquake occurred in the subduction zone of the plate. The main feature of the DONET observation network lies in the dense distribution of testing instruments. The entire DONET observation network is concentrated in the sea area south of the Kii Peninsula; it has a total of 5 nodes and 20 observation points. The distance between observation points is 15–20 km, each equipped with a variety of observation instruments, such as seismographs and pressure gauges. The network can accurately observe various degrees of earthquakes, tsunamis and sea plate deformations.

### 1.5 IORS

The Ieodo Ocean Research Station (IORS) is equipped with 33 types of observation instruments, including 11 types of meteorological instruments, 14 types of marine instruments, 4 types of safety monitoring instruments and 4 types of environmental monitoring instruments. Its main scientific functions and research work currently include comprehensive observations of the oceans and atmosphere, observations of regional ocean characteristics and changes in the Earth's environment, and observations of typhoons and severe weather. Through these long-term observations, key scientific information and data are provided for global change.

## 2. Issues of Recent Underwater Networks

The IoT is the integration of humans, processes and technology with connectable devices or sensors to enable the remote monitoring, manipulation and evaluation of the trends of such devices. The Internet of Underwater Things (IoUT) is a worldwide network for smart interconnected underwater objects that enables the monitoring of vast unexplored water areas [2]. The Internet of Underwater Things is similar to the land-based IoTs in its structure and functions.

### 2.1 Communication Technologies

Underwater telecommunication is difficult, but researchers recently made a huge breakthrough. The research of new underwater wireless communication technology has played the most important role in the exploration of the ocean. Compared to terrestrial radio communications, communication channels in underwater wireless networks can be severely affected by noise, limited bandwidth and power resources, and harsh underwater environmental conditions. For underwater networks, acoustic and optical communications are two types. The present communication technology uses acoustic waves and optical fibers for underwater communication and has low bandwidth, high transmission losses, time-varying multipath propagation, and high latency. The current available acoustic communication technology can transfer data up to tens of kbps over long distances (from 10 m to 10 km). In 2015, Demirors et al. [3] proposed a software-defined underwater acoustic network. A software-defined acoustic modem (SDAM) is, at its core, a software-defined radio (SDR) connected with wideband acoustic transducers through power amplifiers/preamplifiers. SDAM exploits the unique capabilities and features of SDR to fulfill the need for flexible, easily reconfigurable IoT networks.

However, many underwater robots and sensors require a communication link with data rates ranging up to tens of Mbps. Traditional acoustic communication cannot meet this demand. Fiber optic or copper cables are used to achieve high data rates. In 2016, Kaushal et al. [4] summarized the underwater optics wireless communication (UOWC) methods by discussing the underwater optical channel, summarizing various UOWC systems, and discussing the challenges of underwater optical communication.

### 2.2 Tracking Technologies

The IoT usually uses Radio Frequency Identification to track objects. However, as discussed above,

radio waves can not be used well in water. Acoustic tags, radio tags and passive integrated transponder tags are widely used in water to track fish and other objects. Many types of sonar, such as short baseline sonar, long baseline sonar and ultrashort baseline sonar, are used for localization and navigation [5]. AI-Rawi et al. [6] proposed the use of forward-scan sonar (Dual-frequency identification sonar: DIDSON) to capture a scene. A Gabor filter is used to enhance the contrast of sonar images and a Kalman filter is used for tracking. Li et al. [7] proposed a framework to analyze multibeam sonar images to detect and track objects in water. This method used improved Otsu segmentation to detect objects. Lee et al. [10] proposed a simplified real-time color restoration algorithm and then used mean-shift tracking to detect objects. All previously mentioned methods can track underwater objects well in some situations; however, a more practical and applicable underwater tracking framework is needed for real-world use.

**2.3 Energy Harvesting Technologies**

Different from energy harvesting for IoT devices by solar energy and piezoelectric harvesting, ocean current power generation and salinity gradient for electricity generation are the main trends in recent years for ocean energy harvesting. In 2014, IHI engineering [9] reported that they built a floating ocean current turbine system for supplying energy to Ocean of Things networks. The system has two benefits. First, it is a low-cost mooring system that can be used in any ocean. Second, the floating current turbine system can be as the main source of electricity by using ocean current energy that has high capacity with the aid of high-efficiency underwater turbines.

Salinity gradient energy (SGE) is an energy source that was first identified in the 1950s. This energy source relies on energy that dissipates when two solutions with different salinities mix. It is a renewable energy source that is directly linked with Earth's complex water cycle. In this cycle, water evaporates from bodies of water, mainly due to solar radiation. Of course, there are also buoys and underwater surface equipment that use solar energy and novel thermal recharging engines powered by temperature differences.

**2.4 Network Density**

In the Ocean of Things (OT), a very large number of devices communicate to form a network. The critical infrastructure of the OT contains an underwater energy system and underwater communication system. Because of costs and challenges, the OT is sparser than the on-land IoT. Therefore, the distribution of the network is not uniform. The underwater vehicles interact and harmonize through acoustic communication with the network, using sensors to detect and track phenomena of interest. The underlying network architecture of neither is static, as it includes a sensor node for communication and moves a node in the form of the underwater vehicles.

In [10], Mitra et al. proposed a space structured method for active ocean acoustic observation systems using communication constraints. The underwater vehicles interact and harmonize through acoustic communication with the network of the sensor to sense and chase phenomena of interest.

The underlying network architecture of neither was static, included a sensor node for communication and moved a node in the form of the underwater vehicle. In recent years, some optical wireless communication networks [11] have been studied to overcome the limitations of acoustic networks.

## 2.5 Localization Technologies

In the Ocean of Things, localization is important in many applications, and it is different than in the IoT, which can use GPS systems or artificial arrangement to local unknown nodes. Many localization algorithms are being suggested for ocean networks [12]. Research classifies these localization algorithms in two categories: distributed localization algorithms that determine the position of an unknown node and centralized localization algorithms where each unknown underwater node attracts localization information and individually performs place evaluation. In the centralized localization algorithm, the position of each unknown node is estimated by a base station or a sink node. These two categories are divided into estimation-based and prediction-based algorithms. Prediction-based algorithms aim to predict the position of a node using current place information, while evaluation-based algorithms use the current information to calculate the position of the node.

## 3. Cognitive Ocean Network (CONet)

### 3.1 Definition

The proposed cognitive ocean of things architecture is shown in Figure 1. It is divided into four layers: Perception Sensor (Edge) Layer, Local Processing (Fog) Layer, Cloud-Computing Layer, and Application Layer. Each layer uses cognitive artificial intelligence to sense or compute the data. The basic devices and functionalities of each layer are described as follows.

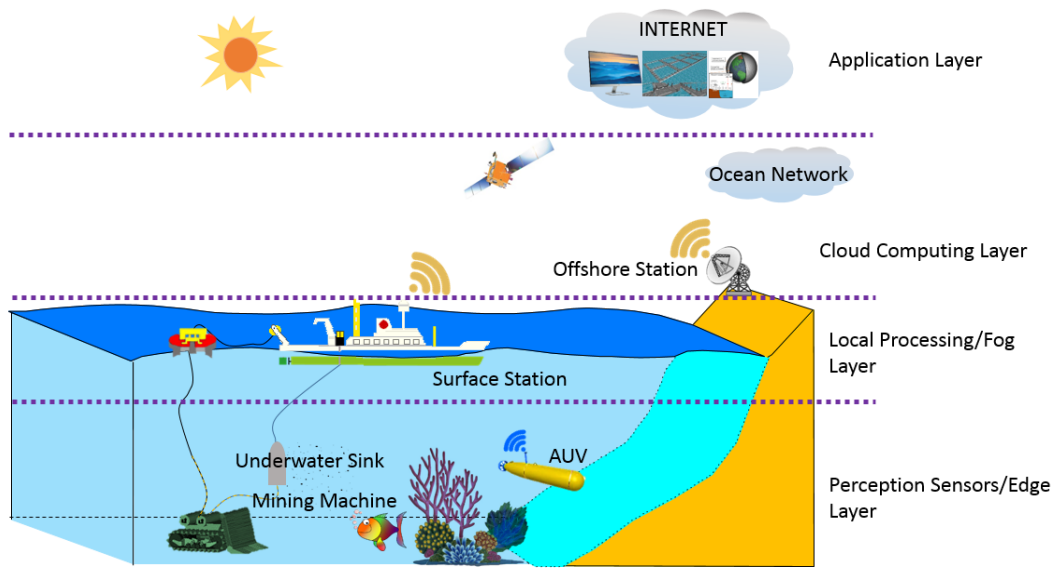

Figure 1. Proposed Cognitive Ocean Network.

### 3.2 Design of CONet

**3.2.1 Perception Sensor (Edge) Layer**

The perception sensor layer, or edge layer, contains various types of underwater sensors, such as automatic underwater vehicles (AUVs), remote-operated vehicles (ROVs), buoys, ships, and automatic surface vehicles (ASVs).

In our proposed CoNet, we develop a motor anomaly detection system to prevent errors of underwater vehicles from operating by measuring the abnormal temperatures. In this system, DS18B20 sensors are attached with the motors to monitor the status of the vehicles. Then, we use deep reinforcement learning, operating abnormalities of the motor are determined by a Raspberry Pi processing unit. The processed signal of control unit is transferred to communication unit and then connecting with the land-based control center through underwater optical communication channels. If the temperature of the motor exceeds the automatically generated threshold temperature that calculated by deep reinforcement learning, the underwater vehicles will return back to the water surface. (see Figure 2).

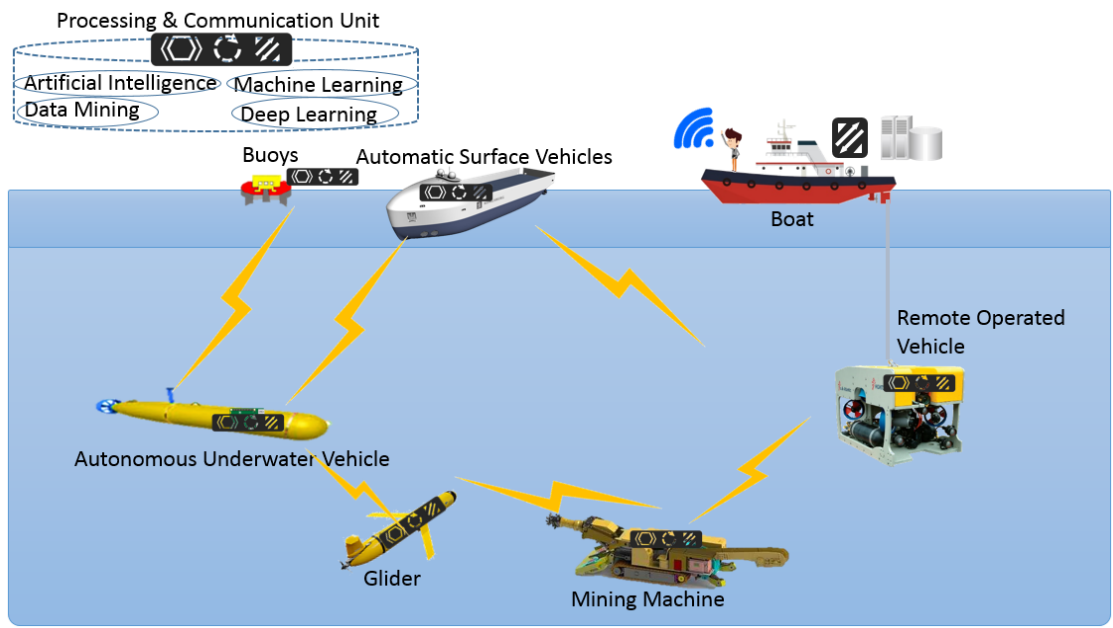

Figure 2. Edge processing using deep reinforcement learning in the CONet system.

**3.2.2 Local Processing (Fog) Layer**

The local processing layer, or fog layer, in the intermediate layer is a bridge between the Perception Sensor (Edge) and cloud computing layers. Many data generated by cognitive sensors in the edge layer can be preprocessed in the local processing layer to shorten the burden of the cloud computing layer. Fog computing nodes intelligently compute to the local zone network level and perform the collection, interaction, analysis and computing of the captured information from many edge devices. Moreover, the local processing layer can protect some security information by using security and privacy algorithms.

In the COnet, we proposed a deep learning-based image super-resolution and descattering method. In the first stage, the underwater camera captured images are descattered by deep neural networks. Then, the input image and the recovered image are super-resolved by traditional super-resolution method. Finally, the fusion scheme is used for selecting the best constraints to output a clear image (see Figure 3).

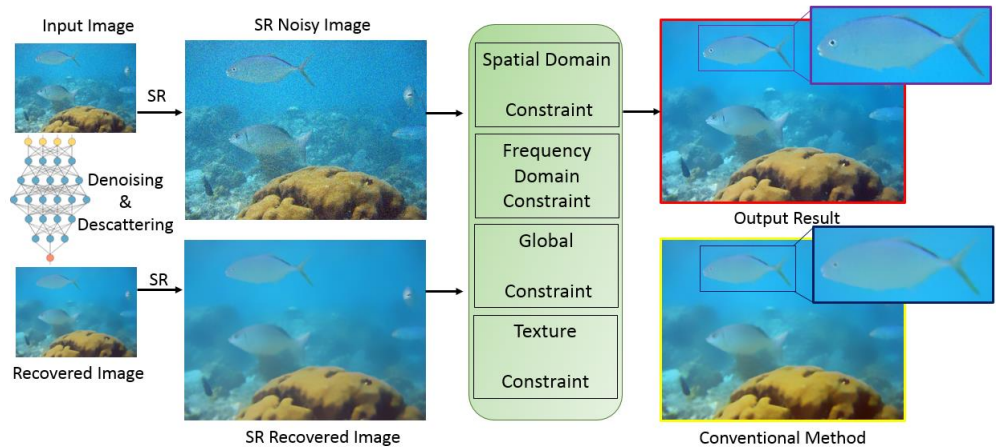

Figure 3. Cognitive descattering and superresolution in the CONet.

### 3.2.3 Cloud-Computing Layer

Information from the local processing layer and perception sensor layer is transmitted to the cloud-computing layer and the related results are sent to the users. This information is retransmitted to the onshore monitoring center using satellite communications, general packet radio service or wideband code division multiple access. In 2016, P. McGillivary et al. proposed ship-based cloud computing for oceanographic research [14]. To shorten feedback between observations and modeling and provide researchers with better visibility into observed processes through data visualization and dissemination via Cloud Services, Schmidt Ocean Institute (SOI) installed a high-performance computing (HPC) system in January 2015 to create the first research vessel with a supercomputing system freely available to the global oceanographic community.

From 2016, Weng et al. [14] have presented underwater image dehazing parallel code with OpenMP, which was developed from the existing fast sequential version on the cloud computing platform. The aim of this work is to present an analysis of a case study to show the development of parallel scatter removal with practical and efficient use of shared memory multicore servers. The results and experiments on the scatter removal application program are executed on multicore shared memory platforms, and results show that the performance of newly proposed parallel code is promising.

### 3.2.4 Application Layer

The World Wide Web is wildly used for CONet applications. In the top level, the CONet applications can be classified as the ocean exploration, environmental monitoring, security guard, disaster prevention, military, navigation and sports. Natural resources detection is important for ocean exploration. For environmental monitoring, such as water quality assessment, marine life habitat tracking and coral reef monitoring, is the key applications for fish farms. The CONet data is used for alerting the floods, volcano, earthquake and Tsunami. Other disasters, such as oil spill can also be detected by CONet.

We detect marine organisms on deep-sea images using YOLO [15]. This tracker is based on convolutional neural networks (CNN) and makes it possible to detect and track multiple objects within an image quickly. YOLO is designed to enable end-to-end learning; in other words, it is possible to simultaneously propose object regions and predict object labels. YOLO uses a single neural network for an entire image. The network divides the image into many regions and predicts bounding boxes and probabilities for each region. It outputs the weighted bounding boxes using predicted probabilities. The traditional YOLO detection network has 24 convolutional layers and 2 fully connected layers. The model is shown in Figure 3 and we describe the outline of YOLO below.

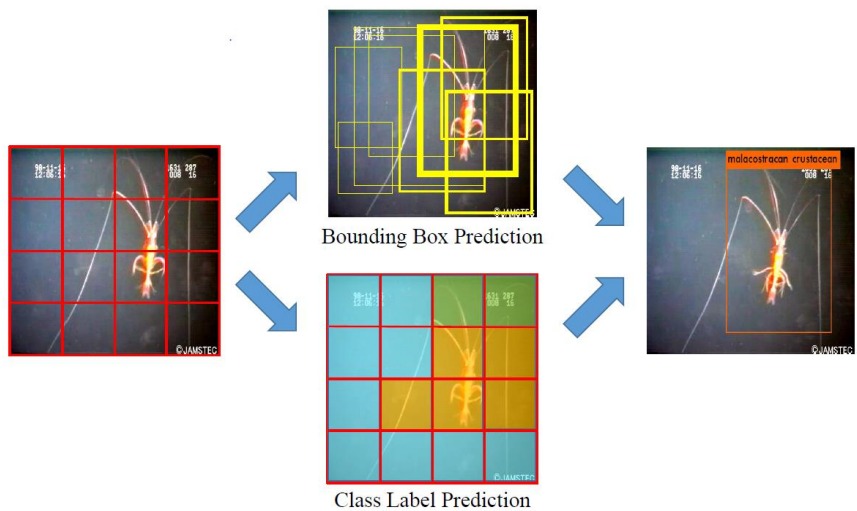

Figure 4. Cognitive tracking model in the CONet system.

YOLO divides an input image into an $S \times S$ grid. If the center of an object falls into a grid cell, that grid cell is responsible for detecting that object. Each grid cell predicts $B$ bounding boxes and the confidence score, $\Pr(\text{Object})$, for those boxes. The confidence score gives the likelihood that an object exists. More precisely, if an object exists in a predicted bounding box, the relative confidence score should be high. Each grid cell also predicts the conditional probability, $\Pr(\text{Class}_i)$, of each class label. A high object label of probability means that the object exists in a bounding box. The confidence score of each bounding box, $P$, is determined by the following equation.

$$P = \max(\Pr(\text{Object}) * \Pr(\text{Class}_i)) \qquad (1)$$

We select bounding boxes whose confidence score $P$ exceeds threshold $T$. In this paper, we use $S = 17$ and $B = 3$.

## 4. Applications

In recent years, many practical and potential CONet applications have been developed. We classify these applications into five categories. The details for each category are described as follows.

### 4.1 Ocean Exploration

Ocean exploration of the ocean is used for many reasons, such as oceanic environment assessment, mineral exploration etc. In Japan, the government is focused on develop new resources from the Ocean. In recent years, deep sea mining is taken in the worldwide. However, there are some problems about ocean mining. First, to find the mines in deep sea is one of the most difficult problems. Second, the technologies for recognizing the mines and mine-like-objects are not so good. To this end, many unmanned vehicles have been investigated for ocean mining. The goal in developing automation for ocean mining is to automatically determine mine location and recognize mine-like objects without humans. There are two stages of ocean mining operations: search-classify-map (SCM), which locates all sufficiently mine-like objects in an operational area, and reacquire-and-identify (RI), which distinguishes mines and processes them accordingly.

With recent operational concepts, automatic underwater vehicle (AUV)-based systems with low-frequency sonar been first used to search-classify-map relatively large areas at a high search rate. Abundant sonar systems, such as side-looking sonar, volumetric sonar, and EO, were developed in the last two decades. SCM is fast but does not determine the mines exactly. Thus, Reacquire-and-Identify is used to finalize classification by reacquiring the target, at a close range, with magnetic, acoustic or electro-optic sensors. Thus, the cognitive ocean of things can solve these problems well.

### 4.2 Security Guard

Security guard often considers the ability to protect from different types of attacks, including underwater attacks. CONet is required for security guard and applied to sea bottom discovery, mineral resource detection and harbor monitoring.

For sea bottom discovery, a sensor and Particle Swarm Optimization (PSO) to measure the water depth to take the position of the sensor nodes. The sensor nodes are placed at various depths and measured by maximizing the surveillance coverage.

### 4.3 Disaster Prevention

Disaster prevention is one of the most important CONet applications because it can save lives and finical lost. An NIED seafloor network along the Japan Trench. A total of 150 seafloor sensor nodes are connected via fiber optical cable. Offshore tsunami sensor nodes successfully caught a 5-meter-high tsunami approximately 10 minutes before its arrival.

**4.4 Other applications**

Other applications, such as using PIT tags to teach visitors fish characteristics, using water quality sensors and underwater cameras for fish farms, controlling and detecting oil spills at deep-sea pipelines, etc., are also the useful for humanity.

**5. Challenges and Future Trends**

Most recent AI models are complicated, overly dependent on big data, and lack a defined function. In this section, rather than merely developing next-generation artificial intelligence technology, we have developed a new concept of general-purpose intelligence cognition technology, called the Cognitive Ocean Network. We summarize the challenges for CONet in five aspects: (1) *self-management*—in the future, researchers must make underwater sensor nodes detect and analyze malfunction by themselves; (2) *energy efficiency*—for long-term long-distance CONet, energy-saving devices must be widely used; (3) *communications coverage*—for sparse CONet, it is necessary to maintain routing automatically (disruption-tolerant techniques that satisfy different application requirements, even in nonstable connectivity); (4) *fog/edge computing*—to make the networking ability robust, fog computing and edge computing should be performed; and (5) *cloud computing*—because of the large amount of data, parallel and distributed computing on the cloud platform should be a main trend in the future.

**6. Conclusions**

In this paper, the Cognitive Ocean Network is introduced. The Internet of Things, artificial intelligence, robotics, and block-chains are core technologies of "Industry 4.0," which will promote the research process of ocean observation. CONet is different from underwater sensor networks (USWs) and the Internet of Underwater Networks (IoUTs) in characteristics and definition, and aims to use artificial intelligence to build intelligent ocean-observing frameworks. The research challenges of CONet have been reviewed and many open problems remain for future investigation.